\documentclass[sigconf]{acmart}
\AtBeginDocument{%
  }

 \renewcommand\footnotetextcopyrightpermission[1]{} 
\copyrightyear{2024}
\acmYear{2024}
\setcopyright{rightsretained}
\acmConference[EAAMO '24]{Equity and Access in Algorithms, Mechanisms, and
Optimization}{October 29--31, 2024}{San Luis Potosi, Mexico}
\acmBooktitle{Equity and Access in Algorithms, Mechanisms, and Optimization
(EAAMO '24), October 29--31, 2024, San Luis Potosi,
Mexico}\acmDOI{10.1145/3689904.3694705}
\acmISBN{979-8-4007-1222-7/24/10}

\begin{document}


\title{Bridging Research and Practice Through Conversation: Reflecting on Our Experience}

\author{Mayra Russo}
\affiliation{%
  \institution{Leibniz University Hannover,
  L3S Research Center}
  \city{Hannover}
  \country{Germany}
  }
\orcid{0000-0001-7080-6331}

\author{Mackenzie Jorgensen}
\affiliation{%
  \institution{Department of Informatics,
  King's College London}
  \city{London}
  \country{United Kingdom}
  }
\orcid{0009-0001-1274-1733}

\author{Kristen M. Scott}
\affiliation{%
\institution{DTAI,
  KU Leuven}
  \city{Leuven}
  \country{Belgium}
  }
\orcid{0000-0002-3920-5017}

\author{Wendy Xu}
\affiliation{%
  \institution{Department of Economics,
  The Chinese University of Hong Kong}
  \city{Hong Kong SAR}
  \country{China}
    } 
\orcid{0009-0002-5683-5296}

\author{Di H. Nguyen}
\affiliation{%
\institution{School of Mechanical and Materials Engineering,
  University College Dublin}
  \city{Belfield, Dublin}
  \country{Ireland}
  }
\orcid{0000-0003-3592-6287}

\author{Jessie Finocchiaro}
\affiliation{%
  \institution{Department of Computer Science,
  Boston College}
  \city{Chestnut Hill}
  \state{MA}
  \country{United States of America}
  }
\orcid{0000-0002-3222-0089}

\author{Matthew Olckers}
\affiliation{%
  \institution{e61 Institute,
  Macquarie University}
  \city{Sydney}
  \country{Australia}
  }
\orcid{0000-0001-7096-7047}

\renewcommand{\shortauthors}{Russo et al.}  

\begin{abstract}
While some research fields have a long history of collaborating with domain experts outside academia, many quantitative researchers do not have natural avenues to meet experts in areas where the research is later deployed. We explain how conversations---interviews without a specific research objective---can bridge research and practice. Using collaborative autoethnography, we reflect on our experience of conducting conversations with practitioners from a range of different backgrounds, including refugee rights, conservation, addiction counseling, and municipal data science. Despite these varied backgrounds, common lessons emerged, including the importance of valuing the knowledge of experts, recognizing that academic research and practice have differing objectives and timelines, understanding the limits of quantification, and avoiding data extractivism. We consider the impact of these conversations on our work, the potential roles we can serve as researchers, and the challenges we anticipate as we move forward in these collaborations.
\end{abstract}



\maketitle

\section{Introduction}

\label{sec:intro}

Many researchers trained in quantitative methods center their work around a social issue but have limited to no connection with experts who work on the issue on a daily basis. For example, researchers in fair machine learning tackle domains like criminal justice and finance but often fail to incorporate specific domain knowledge and the historical and social contexts surrounding these issues~\citep{selbst2019FairnessAbstractionSociotechnical,katell2020SituatedInterventionsAlgorithmic}.

Certain research fields present more opportunities to interact with practitioners than others. For example, development economists partner with non-profit organizations to engage in fieldwork and impact evaluations of intervention programs~\citep{banerjee2009experimental}. Community engagement and public involvement are common in public health, where interactions between researchers, healthcare providers, and affected communities directly guide research directions and inform research findings~\citep{omara-eves2015effectiveness}. While some quantitative researchers are part of projects or departments where researcher-practitioner partnerships are common practice, the vast majority have limited opportunities to interact with practitioners.

To support quantitative researchers seeking to center their research on social issues, we present our approach to foster conversations with practitioners. We define conversations as interviews without a specific research agenda or expectations from any party for future collaboration. This approach provides opportunities for researchers to connect with practitioners---people who actively work with affected communities or influence related policies---to learn from their expertise and experience. Our approach is especially relevant for researchers without an established avenue to interact with practitioners.

We implemented the conversations approach as part of a working group with a broad focus on addressing the needs of disadvantaged or marginalized groups. Over the past three years, our group has interviewed over 20 different practitioners, including a lawyer, a medical doctor, a former government minister, social workers, and representatives of non-profit organizations. The practitioners were based in many different countries, including Kenya, Australia, Germany, and the United States, and worked on a range of different domains, including refugee rights, conservation, addiction counseling, and municipal data science.

Our experience learning from practitioners has been transformative, prompting us to reflect on the process and its outcomes. Initially, we started writing a blog post summarizing what we learned from the diverse range of practitioners we interviewed over the past three years. That effort grew into a desire to more fully define, critique, and reflect on our conversations approach. We identified a need to synthesize what we learned and to delineate how our perspectives shifted through these conversations. Thus, we moved from writing a blog post to reflecting on our ideas more formally in this paper. 

We have used the methodology of collaborative autoethnography to reflect on our experiences. We engaged in this process to answer the question: How have quantitative researchers looking to address the needs of marginalized groups experienced conversations with practitioners? Throughout this paper, we present our own experiences from the conversations and the process. 

Based on our experience, we argue that: 
    \begin{enumerate}
        \item  While practitioners are often in need of resources and support, researchers need to be willing to adjust to the practitioners' priorities and embrace their unique understanding of the context. We learned to value the knowledge of experts, embrace the differing objectives of research and practice, avoid data extractavism, and understand the limits of quantification.
       \item Our outlook on our work has shifted greatly. We have adjusted our expectations on the time it takes to build relationships and develop relevant research projects with practitioners, we have an increased familiarity with and interest in qualitative methods, and we have a more realistic idea of the challenges researchers face in interacting with practitioners.
    \end{enumerate}

We conclude by proposing activities for researchers to build on their connections to practitioners. These activities include but are not limited to providing administrative tech support as volunteers, translating the jargon of research and practice, educating students on real-world issues with case studies from practitioners, and advocating for systematic reforms to better serve marginalized communities.

\section{Defining conversations with practitioners}

In this section, we define our conversations approach and discuss how conversations relate to different types of interviews. We explain how we implemented our conversations with practitioners as part of a working group. 

\subsection{Definition}
We define conversations as interviews without a specific research agenda or expectations from any party for future collaboration. Interviews allow people to share their stories and experiences fully and in their own words; this is particularly important if the priority is seeking to understand other perspectives~\citep{creswell2003research}. The intended result is not a set of numerical values statistically analyzed as in quantitative methods. Generally, interview transcripts, recordings, or notes are analyzed closely and iteratively with the purpose of understanding and synthesizing the meaning in relation to the researchers' specific questions~\citep{braun2006TA}.

Interviews are categorized according to different criteria. One criterion we can use is whether the interviews are structured, semi-structured, unstructured, or informal interviews~\citep{gubrium2002handbook}, depending on how strictly the interview is guided by a pre-determined list of questions. Most of our conversations take the form of semi-structured interviews and include a significant amount of unstructured interview time, in which all present can participate. Another categorization of interviews is according to their roles in a research project or pipeline. ~\citet{starr2014qualitative} categorizes interviews into exploratory and explanatory, where exploratory interviews aim to frame research questions while explanatory interviews aim to falsify hypotheses. Within this dichotomy, our conversations are exploratory interviews.

\subsection{Our implementation}
To organize our conversations with practitioners, we created a working group. Each member of the working group took a turn once a year to lead an interview with a practitioner of their choosing. Although conversations do not need to be organized through a working group, we found the peer support helpful. 

When a group member had their turn to lead an interview, they took charge of identifying a practitioner of interest and inviting them to the interview. For the most part, we sought out individuals outside our typical research work. For example, one of our members interviewed a social worker who counsels refugees even though this member had never conducted research related to refugees.

We did not use any formal approval process for choosing who to interview, and we did not have any restrictions on the practitioners our group members could choose to interview. Each group member had the freedom to choose a practitioner they were interested in learning from. Our group members have a diverse range of research interests, so allowing the freedom to choose any practitioner worked well for us. Groups with a shared research interest may benefit from focusing only on interviewing practitioners from a specific domain.

The lead interviewer starts the conversation by first allowing the practitioner to introduce themselves and their organization and to give an overview of their daily operations. Following this introduction phase, the lead follows with a set of questions that they prepare in advance. However, the aim is to have group members interact with the practitioner throughout the conversation---we always provide an opportunity for unstructured discussion between the practitioner and all present. We hope this spontaneity can create inclusive spaces and leave room for practitioners' own descriptions of their social contexts, as well as encourage participating researchers' curiosity, especially about topics outside their expertise.

After each interview, we summarize the conversations and the practitioners' insights in a blog post shared through social media. We choose to disseminate and communicate what we take away from the conversations via blog posts because of their suitability for sharing stories with a broad community of researchers across fields. Blog posts can also be prepared more quickly than qualitative academic research articles. The lead interviewer prepares the blog post independently and receives feedback from other group members and the interviewed practitioner. We highlight that we maintain the practitioners' anonymity throughout this paper since our agreement with them was only to summarize their interviews in blog posts where they had editorial control over the content.

Blog posts are also a more accessible output format for quantitative researchers to report on interviews than conducting a rigorous analysis using qualitative methodology. Even leading researchers in qualitative methodology concede that qualitative methodology can be overwhelming. In their book on thematic analysis, \citet{braun2022TA_book} explain that: "One of the deeply confusing elements is that qualitative research can be underpinned by very different, and sometimes contradictory, theoretical ideas. Many qualitative scholars like to write about these things---and often in ways that seem like a spiraling tornado of ideas and debates, fast-moving, impenetrable, and somewhat terrifying." Our blog posts provided an accessible entry point to interact with qualitative data.

\subsection{Connection to similar approaches}

The conversations with practitioners approach can be considered as part of Action Research, a set of research methodologies that requires collaboration between researchers and practitioners to promote social changes~\citep{checkland2007action,greenwood2006introduction}. Different variations of Action Research have been developed to address an agenda driven by the practitioners versus by the researchers~\citep{hammersley2004action}. For instance, Participatory Action Research emphasizes the involvement of community members affected by the research; these members are treated as co-researchers~\citep{mcintyre2007participatory}. Community-Engaged Research is the process of working collaboratively with groups of people affiliated by geographic proximity, special interests, or similar situations with respect to issues affecting their well-being~\citep{yalemed_CER}. A significant task of this research involves defining "the community" who may not share the same experiences or opinions. One of the most widely used forms of Community-Engaged Research is Community-Based Participatory Research, which is widely adopted in health research~\citep{wallerstein2020engage}. 

While called different names, the aforementioned research approaches center on the value of lived experiences and the context from where issues arise to drive research and solve challenges. Action Research variations address the specific needs of the applied field and balance between research and practice. Increasingly, social activists and organizations advocate for more responsible approaches when applying technology in quantitative fields to aid in solving societal problems. Conversations with practitioners aim to provide a low-barrier entry for academic researchers to learn more about the problems they seek to study.

\section{Collaborative Autoethnography}

\label{sec:autoeth}

To reflect on our experience as quantitative researchers attempting to learn from practitioners, we use the qualitative research method of collaborative autoethnography \citep{chang2013}. The method can be defined by breaking up the name into three parts: (1) collaborative, (2) auto, and (3) ethnography. We will explain the parts in reverse order. Ethnography is the study of human groups and cultures by immersing oneself in the group. Typically, ethnographers aim to take an objective and detached view of the group they study. The `auto' in autoethnography flips this traditional approach. Rather than studying other people, the autoethnographer reflects on their own experiences to provide insight into a culture or society. Finally, the collaborative in collaborative autoethnography is to tackle the process of self-reflection as a group. Bringing all three parts together, collaborative autoethnography involves a team of researchers who share and analyze their personal experiences related to a particular cultural phenomenon or context. In this paper, our team of researchers shared and analyzed our personal experiences of holding conversations with practitioners to bring insight into the gaps between quantitative research and practice.
 
We collected data on our experiences by each separately answering questions on what brought us to the group, including our thoughts on interviews as a method and our reflections on what we learned from practitioners (see Appendix \ref{sec:appendix} for the questions). We analyzed our answers by finding patterns and key insights through an iterative process of collective reading, theme identification, discussions, and writing. Throughout the process, we reviewed existing literature that related to the topics that we discussed. The writing of our collaborative autoethnography was organized through regular online meetings. We have made frequent use of quotes to highlight members' views and experiences. Our approach to analysis is closest to the method of reflexive thematic analysis \cite{braun2022TA_book,braun2006TA}, which embraces the researcher's perspective rather than trying to separate the researcher's biases and assumptions. 

We did not set out to conduct a collaborative autoethnography at the start of this project. We initially planned to write a blog post together that summarized what we had learned from the diverse range of practitioners our working group had interviewed. During the process of preparing the blog post, we aspired to conduct a deeper analysis to reflect on our conversations approach, so the blog post evolved into this paper. We chose the method of autoethnography as we wanted to examine the impact the conversations had had on us as researchers. Our starting point of synthesizing solely what the practitioners had shared with us was not satisfactory because, first, it was insufficiently introspective and self-critical, and second, we did not want to (and did not have consent to) take the approach of studying our interviewees as research subjects.\footnote{We asked our interviewees for consent to publish summary blog posts about their interviews but did not ask for consent to include their interviews in academic studies.}

The autoethnographic viewpoint allows us to turn our research gaze inwards, making us simultaneously the "researcher and the researched''~\citep{chang2013}. In this way, we can examine, through our own experiences, the phenomena of researchers interacting with practitioners and engaging in unfamiliar research methods.

\section{Reflecting on the process}

In this section, we share our motivations for joining the conversations working group and reflect on the process of conducting interviews with practitioners. We discuss both the challenges we faced, such as initial hesitation in approaching practitioners and concerns about using appropriate language, as well as the successes we experienced, including hearing diverse perspectives and building new connections.

\subsection{The group and our motivations}
\paragraph{Positionality Statement}
We acknowledge that our research, findings, and takeaways are shaped by the lived experiences of our research team. The group hosting practitioners for interviews generally have technical backgrounds in computer science, economics, industrial engineering, operations research, and information science. All authors have been formally trained in quantitative research methods. A few of us have experience using qualitative methods but are not formally trained. Our team is currently distributed across North America, Europe, Asia, and Australia; we acknowledge that our locations affected the selection of the practitioners with whom we connected. We recognize that while we include discussions of the needs of marginalized groups, we have specifically held interviews with practitioners. We treat practitioners here as representatives of community-based organizations~\cite{costanza-chock2020DesignJusticeCommunityLed} and as a contact point with marginalized communities.

\paragraph{Motivations}

In our autoethnographic exercise, we asked ourselves to think about our original motivations and expectations for joining our working group. Many of us had a strong desire to center and amplify marginalized voices in our work but also had misgivings about entering a space in which we did not have a background. As
Jessie shared:\begin{quote}
    \textit{"Before I proceed in this space, I want to learn more about other areas and learn how practitioners think so that, if there is a helpful algorithmic model for a problem of their interest, we are not algorithmicizing the problem in a harmful way."}
\end{quote}
We were also drawn to an opportunity to expand our horizons in this area, which Wendy highlighted: \begin{quote}
    \textit{"I was excited to see STEM researchers gather for conversations and interaction with practitioners. This shows respect for the value of domain expertise working directly with marginalized communities."}
\end{quote}

Despite its importance, we note that the contextual understanding of social issues is often not highly valued in quantitative research. As Mackenzie noted, \textit{"it is easy for academic research to lose touch with the real world and implementation, ...[I] joined the group as a way to counter academic complacency."} 
In a similar light, but with a different perspective, Matthew mentioned how a big motivator to join conversations was to fill a gap he felt existed in his training as a quantitative researcher: \textit{"even though I was trained to analyze survey data, the survey questions were only numerical or categorical. My training avoided how to analyze open-ended questions."}

\subsection{Reflections on the conversations process}
\paragraph{Challenges}

In addressing the challenges emerging from the process of conducting interviews, we noticed a dichotomy. Most of us found that organizing an interview and participating in this interactive knowledge exchange felt like a fruitful way to spend time with our peers. However, quite a few of our members did not know where to start, and cold-calling practitioners to speak with us felt daunting. In a few instances, some of us managed to sidestep this by resorting to our personal networks; Mackenzie \textit{"found having a personal relationship with an individual at an organization or a connection of mine who could introduce me led to responses."} However, this was not an option for most of us; thus, we felt pressured to make sure we were prepared when approaching practitioners outside our networks. Jessie shared that to make a positive first impression, \textit{"the most helpful thing has been to try to do our homework beforehand."}

When first contacting practitioners, we found that carefully crafting our email messages was crucial. Our goal was to persuade busy professionals to volunteer an hour of their time for our project. While thorough preparation helped alleviate some concerns, it didn't completely eliminate them. As Di pointed out, we worried about inadvertently \textit{"using demeaning/marginalizing language (i.e., saying things that can be considered social norms but are actually harmful to the community in question)."} We acknowledge that language conveys a perception of group identity, which can be subject to longstanding social suppression and hidden discrimination. This worry conveys a specific challenge of lacking experience with interactions with practitioners or marginalized communities.

When we did receive a positive answer to our inquiries, we then dealt with another set of minor challenges. These included, for example, keeping in contact with the practitioners and adapting to their schedule changes. If the practitioners needed to shift the interview date, it was usually due to their busy schedule and the reality of attending to unexpected urgent work. As noted by Mayra,
\begin{quote}
    \textit{"Another concern I had arose from scheduling an interview two months in advance. This brought up a lot of anxieties in general, as life often gets busy, and I wasn't sure if the expert was going to fall through for any reason."} 
\end{quote}

\paragraph{Successes}
When reflecting on our conversations, we collectively identified the key success of hearing from a diverse set of voices in our conversations. Our practitioners' expertise spanned a variety of domains with common themes, including municipal data science, human rights, immigration, and conservation.
The interviewees varied greatly in their roles within those domains (e.g., lawyers, social workers, data scientists, and advocates).

In addition, several of us reflected on how arranging and partaking in the interviews provided us with a way to build new connections. Kristen emphasized the importance of maintaining relationships with practitioners after the interviews so that: \textit{"they did not feel used, and we also made sure they were on board for reviewing the eventual blog post."} Through this process, we developed our relationship-building skills. As Kristen observed: \textit{"there is a social aspect, a relationship-building aspect that is its own skill and not everyone's forte."}

As a group, we also perceived successes in the way that these interviews met our initial expectations upon first joining the group and served as reality checks. Specifically, Di expressed:
\textit{"Research takes time. Realistically, there is no reason for me to believe I could, in a few months or years, immediately do research that can positively impact the community."} 
Moreover, we also appreciated how our own knowledge was often challenged and how we learned more about what we did not know. In this process, we grew as more informed and knowledgeable researchers and people, as noted by Mayra:
\begin{quote}
     \textit{"The conversations create a space where we can learn about the tensions, the creases and also as a way to recalibrate our expectations as to what it is to work in projects with direct impact in society and the compromises that must be made, etc."} \end{quote}

\section{Lessons learned}

\label{sec:lessons-learned-from-pract}  

In this section, we discuss recurring themes brought up in many of our conversations with practitioners. We highlight four lessons. Researchers should (1) value the knowledge of experts in social contexts, (2) embrace the differing objectives of research and practice, (3) avoid data extractivism,  and (4) understand the limits of quantification.

\subsection{Value the knowledge of experts}

Our interviews revealed how academic researchers often approach technical interventions, such as algorithmic systems, as novel solutions divorced from historical context. This perspective can inadvertently perpetuate existing inequities and injustices~\citep{hoffmann2019fairness,weinberg2022rethinking,joyce2021toward,birhane2021algorithmic}. As Di explains, the interviews often challenged our preconceived notions:
\begin{quote} \textit{
    "Conversations with practitioners have also introduced me to new ideas and challenged some of the social conventions that I have accepted. I think it is good for me as a researcher to think about the implications these have on my work and vice versa."
} \end{quote}
For example, Di gained a new perspective on substance use disorders, which are often stigmatized as being caused by a lack of self-control. Through her interviews, she gained the perspective that substance use disorders are a medical need that should be treated as such, without the stigma.

The practitioners we spoke to emphasized the importance of listening to affected people and on-the-ground experts. 
They discussed how citizens want to influence policy issues that affect their day-to-day lives. 
Continuous dialogue with marginalized communities is crucial because of the evolving need for policy interventions~\citep{atkinson2021dynamics, zaller1992nature, noelle1993spiral}.
Similarly, our conversations highlighted that any successful technology or tool will require extensive contextual knowledge. 

Researchers moving from academia to the non-profit sector, in any capacity, should be patient when getting to know the sector and learning to detect needs from the bottom up~\citep{babajanian2005bottom,wessells2015bottom,smith2008critical}. These researchers will find that real problems are too complex to easily quantify and model. For this reason, practitioners expressed the need to listen to those affected and learn from their lived experiences directly. 
In practical terms, and to encourage knowledge exchange, some of the practitioners we talked to host and organize sessions that focus on the active participation of affected people and discussion of the different issues present in their communities. From these sessions, practitioners aim to better understand what levers of change are needed and possible in order to develop stronger strategies, advocacy, and research. In our exchanges, practitioners expressed that they do not aim to extract expertise from people; rather, they aim to work with them and contribute in a productive and socially-aware way~\citep{fisher2015mapping,bouwen2004multi}. 

While learning from domain experts is valuable, we also became cognizant of the need to learn from and converse with a variety of experts. Multiple practitioners within one domain likely have different experiences---they potentially even hold different viewpoints and perspectives on problems within their domain. We must listen to opposing voices. The practitioners we interviewed also recognized this; they cannot work, research, nor advocate in isolation~\citep{fung2015putting, reddel2004consultation,shevlin2022respecting}. While they are experts in their fields and in specific topics, the practitioners acknowledge that they are not experts in all matters with respect to their work. For this reason, they value carefully crafted collaborations with researchers, as they often lead to mutually beneficial cross-pollination.

\subsection{Embrace the differing objectives of research and practice}

Practitioners and academic researchers approach their work with different objectives and incentives. For instance, academic work must emphasize its novelty in comparison to existing contributions, making projects span months or years. In our conversations, many practitioners discussed the urgency of their work. Rather than months or years, practitioners must tackle immediate problems in adequate ways within days or weeks. While seemingly innocuous, these differences reveal that practice and academic research can take opposite stances on the trade-off between innovation and speed. As a result, these frictions can create challenges in researcher-practitioner collaborations. 

In the context of the education sector,~\citet{Coburn2013-pi} note that "academic research rewards novelty and precision, whereas the practitioners require action and information in a fairly timely way." 
While some day-to-day activities may move quickly in the practitioners' world, there is often a longer timeline to their high-level organizational work. This is due to the fact that their work often addresses longstanding and chronic issues. 
On longer timelines, practitioners dedicate crucial time to relationship-building within the communities in which they work and to learning about the unique context in which they are engaging. 

From our conversations, we learned that for researchers collaborating with practitioners, there must be flexibility in terms of project planning. While hitting milestones is important, we learned that practitioners prioritize that relationships are built, people understand what is happening, and some positive outcomes related to the organization's goals are achieved~\citep{hardina2021strategies, SALINGER2024106449}. ~\citet{le2015strangers}
emphasize that in researcher-practitioner collaborations, the plurality of voices within the community where the practitioners' work is set necessitates a prioritization of interpersonal relationships. Further, practitioners place value in personal interactions, especially informal ones, since they help build trust, develop sustainable community engagement, and form allyship with other organizations in advocacy activities~\citep{w1997community,whittaker1983social,coupland2003small,vangen2003nurturing}.
 
\citet{Coburn2013-pi} highlight how the execution of certain collaborations can dissuade researchers from actively participating in these collaborations because of how their productivity and research output are measured and evaluated at their academic institutions. This is a reality that is ever present for researchers, as one of our team stated: 
\begin{quote} \textit{"Some efforts devoted to the collaboration are not identified as research output. Considering the knowledge burdens in academic publications are high and still growing, insufficient recognition [by institutions] of researchers' efforts in the researcher-practitioner collaboration deters researchers' sustainable engagement."}
\end{quote}

The trade-off between local relevance and researchers' pursuit of generality imposes challenges to these collaborations and tests their capacity to address those cases at the margins. ~\citet{d2010center} observes this phenomenon in education, highlighting that "as [researchers] work with larger numbers of schools or districts, they no longer have the capacity to truly be responsive to local issues and multiple entities with diverse needs." 
Additional challenges arise from the overhead incurred in designing and developing intervention programs together with individual organizations~\citep{Krajcik2010-df, Roderick_2009}. To address these challenges, researchers would have to give significant time, and this work is often not credited by their institutions. ~\citet{Krajcik2010-df} observe that "universities do not provide sufficient support for researchers spending time on the collaborative work with practitioners or engaging in efforts to provide information to the public."

Researchers seeking to collaborate with practitioners must understand the differing objectives at play and embrace them. They need to position themselves to align their work with practitioners and the community needs, even if that means slowing down their institutionalized research agendas, reminiscent of the slow-science philosophy~\citep{stengers2016another}.
In conversing with practitioners, we created a channel that vicariously allows us to start \textit{"thinking, connecting, and working with the community,"} as Di shared. We can prioritize the short and long-term effects of our own research by better understanding the problems that practitioners face.

\subsection{Avoid data-extractivism}

Data collection is a practice that is deeply embedded across academic research. 
However, the move towards a "digitized society" has put more pressure on the need to acquire data at any cost. For example, we can see this concern reflected in the current public discourse surrounding the data used to train large-scale AI models, such as those requiring billions of words or hundreds of millions of images---often scraped from the internet without clear consent.\footnote{\url{https://www.adalovelaceinstitute.org/resource/foundation-models-explainer/}} This practice raises questions about data requirements, provenance, and ownership.

Oftentimes, researchers, corporations, and governments, among others, collect their data by foregoing ethical and data minimization principles designed to protect the dignity and privacy of individuals~\citep{doi:10.1177/20539517231206802,10.1145/3630106.3658543}. For example, reports have claimed that technical companies hired contractors to target low-income individuals, such as chronically homeless people and college students; instructed to resort to deceptive practices if needed, they would offer individuals financial compensation in exchange for photos of their faces in an effort to diversify facial recognition datasets that improve facial recognition technology embedded in their systems.\footnote{\url{https://www.cnbc.com/2019/10/03/google-contractor-reportedly-tricked-homeless-people-into-face-scans.html}} Similarly, researchers on platform work have uncovered how technical companies recruit gig workers to sell their personal data (e.g., selfies, government-issued identification documents, voice notes) to produce AI training datasets for very little compensation and in the context of murky or non-existent regulation that can protect these individuals from being subjected to exploitation \cite{AW_P}.

The increased data dependency, reinforced by the emergence of data-intensive tools and applications~\citep{doi:10.1177/20539517231206802}, can cause us to partake in these harmful practices.
Furthermore, when viewed through a critical lens, these data collection practices rehash historical extractive practices, creating and maintaining significant socioeconomic inequalities and disparities~\citep{resistingcol,crawford_atlas_2021,10.1145/3630106.3658543}. 
From our reflections, Mackenzie conveyed that practitioners highlighted this concern and expressed what researchers should work towards instead: 
\begin{quote}  \textit{"I learned that data extractivism is a major issue, which is the phenomena of researchers or more powerful groups taking information or resources from marginalized communities and not working with them or giving their support in return. Data extractivism has colonialist underpinnings, so we should aim to work with these groups and use more participatory research methods." } \end{quote}

Researchers working with vulnerable communities should resist data extractivism despite potential incentives or perceived inevitability. However, limited resources often create tensions between stakeholders, requiring careful consideration of trade-offs~\citep{whittlestone2019role, mittelstadt2019principles}. Our conversations revealed several such tensions, including balancing data access and open-data initiatives with the risk of overexposing vulnerable communities. Researchers must proactively address these complex challenges to mitigate potential sociotechnical harms~\cite{borgesius2015open, weller2019transparency, winter2015big, duncan2004exploring, filip2022regulating, 10.1145/3600211.3604673, 10.1145/3630106.3658543}.

Finally, our conversations also highlighted the importance of closing the loop and communicating information for data literacy. When researchers and practitioners work on community-driven projects, they must implement grounded dialogue models of communication and interpretation~\citep{madumal2018towards,remington2010communicating}. With regard to AI in particular,~\citet{baroni2022ai, zheng2022putting, rahwan2018society}, all advocate for the comprehensive engagement and empowerment of affected persons across the AI project pipeline, including in data collection. 
They argue that equipping people with the needed understanding helps reject the narrative that algorithms are incomprehensible black boxes that people cannot relate to nor interrogate. One step beyond that, from our conversations, we learned that promoting awareness activities and engaging in dialogue with affected communities can highlight simpler and more straightforward methods to intervene (than highly complex and opaque models)~\citep{rudin2019stop, rodu2020black,brozek2023black}.

\subsection{Understand the limits of quantification}

Limited human and financial resources make automation through technological development tempting. 
However, from our conversations, we learned that an excessive focus on data-driven processes, automation, and scaling up can become antithetical to the goals of organizations working on community-based projects. Goals often revolve around listening to affected voices, advocating for policy changes, and community enrichment support. 

Practitioners highlighted how over-relying on data-driven decision-making to inform policymaking commonly leads to proposed solutions that only superficially address societal problems. 
We emphasize how co-creation and collaboration with marginalized communities are necessary to bring the root underlying issues to light, helping us understand inevitable gaps in the data~\citep{costanza-chock2020DesignJusticeCommunityLed}. 
~\citet{d2023data} demonstrate how direct contact and interaction with local communities shapes data analysis, particularly situating data within contexts and incorporating cultural and social factors.

There is a high risk that solely technological or data-driven-based interventions will fail to address, and may even detract from, the priorities of practitioners. ~\citet{burton2009accountable} discuss the pitfalls of solely data-driven or technology-based solutions in social service delivery and the social work sector. 
Technological interventions that strip frontline workers of autonomy intensify the tensions between management and frontline social workers. Managers in charge of technology initiatives often gravitate towards their implementation, dictating data collection protocols and deciding how the data will be used. 
~\citet{reardon2010data} discusses operational, philosophical, and financial reasons for the negative effect of information management systems and provides several examples of how "the frontline workers often find the information systems to be of limited value." ~\citet{huesemann2003limits} demonstrates a similar view in sustainable development that technology-driven solutions are insufficient for sustainable development goals. 

Another recurring challenge mentioned by our interviewees is the frustration associated with quantifying success. 
Almost all organizations face pressure to measure their social impact, making it difficult to escape the confinement of formal measurements. 
The importance of advocating for multiple ways to measure impact became evident. 
Some of the alternatives included recognizing informal work and interventions from day-to-day activities that help and contribute to building relationships and trust with local stakeholders~\citep{falk2000social,stoll2010informal,lambright2010building,delgado1999community}. 
Specifically, practitioners need alternative means to assess the value of their work beyond quantification~\citep{garland2013mindfulness,plath2006evidence}.
Interpersonal interaction, for example, is an essential aspect of their work; it cannot be easily quantified in numerical terms~\citep{unrau2007evaluation}.

While there is room for quantitative analysis and algorithmic tools to offer insights, we recognize they are viewed as neither new nor without controversy and limitations by our interviewees. 
Often, the stories we can tell without quantification or technical intervention are more important than any traditional quantitative measure of success. Mackenzie also highlighted that: \textit{"...metrics and quantitative evaluation techniques depend greatly on the context they are used in; there is not a one-size fits all answer."}

\section{The road ahead}

\label{sec:road-forward}
In this section, we reflect on our realities as researchers, moving forward with the lessons described above. We draw heavily on the reflections made in the autoethnographic process described in Section~\ref{sec:autoeth}. In particular, we reflect on the impact our experiences with conversations with practitioners have had on our perspectives on working with marginalized groups, the challenges we foresee in approaching that kind of work, and finally, consider potential roles for researchers such as ourselves when approaching work on sociotechnical problems.  

\subsection{Impact on our work}
We recognized the need to take a long-term perspective when considering how to incorporate practitioners and marginalized groups into our research work. We need to take the time to build relationships and learn about the specific context in which we want to work rather than immediately imposing a solution or approach that works for our own research performance indicators. We assume that building long-term partnerships with organizations on the ground is most beneficial at this stage. Later on, research could grow out of these partnerships, but publishable research is not the primary goal. 
Also, we found that our participation in the group reminds us to use a critical lens in our own work, not only when working with marginalized groups, but also when designing experiments, writing papers, and making claims about proposed methods.

The conversations group has had an impact on what kind of research we see as valuable and even doable. Overall, our shared interest in qualitative and interdisciplinary research has grown. For some of us, exposure to, or training for, qualitative work has been minimal or non-existent. Others have even witnessed a stigmatization of qualitative methods in their fields. One member explains that qualitative methods are \textit{"completely underrated and undervalued in my computer science lab"} and Jessie shares \textit{"I admire qualitative research, but I don't think it will get me tenure when I was hired as a CS theorist."} Matthew recalls that \textit{"Working with the group accelerated my interests in qualitative methods. From my engagement with the group, I started to include more open-ended questions in my surveys."} Mayra also reflects that \textit{"All types of research have their pros and cons, and I think there's a lot of value in incorporating and mixing different types of methodologies [...] I feel like nobody has the sole answer as to how to do science,"} and goes on to say that she hopes to continue to draw from qualitative research methods as well as quantitative in her career. 

We also highlighted the importance of engaging with qualitative data, such as interviews and stories, to properly understand people's experiences and realities. Qualitative research training is often not available to quantitative researchers. Mackenzie suggests that \textit{"students in quantitative degrees should have more options to learn about more qualitative methods."} Through qualitative methods, we can more easily see our research from different angles, which is not only beneficial to us but those who could be impacted by it. 

We found that our engagement with one another not only increased our interest in qualitative methods but also increased our confidence in them. Matthew describes the experience as follows: \textit{"Engaging with qualitative methodology was not as scary as I had imagined. It actually feels like a breath of fresh air to look at research from a completely different angle from what I have been trained to do."} Wendy mentioned getting to know new possibilities for interdisciplinary exploratory research through the option to submit position papers to \textit{"conferences like [EAAMO]"}\footnote{\url{https://conference.eaamo.org}}.

\subsection{Challenges we foresee} 
Given that practitioners work on complex social problems, the issues that they face, as well as the support they seek, are sometimes not particularly related to technology-based solutions. One significant challenge that practitioners and community groups commonly struggle with is funding and resources; Mayra explains that: \textit{"all resources are scarce and in the third sector especially you are more prone to work more for less in terms of personal retribution"}. Resource limitations often include technological resources, such as software products, infrastructure, and tech know-how. As Jessie states, \textit{"Most of our conversations have revealed that digitization is not available for many marginalized folks."} As researchers trained primarily in advancing scientific knowledge, we find ourselves at an impasse when presented with the realities and stated needs of the practitioners with whom we hope to collaborate.

Beyond feeling that our efforts might not be particularly useful, we are also concerned that our work with practitioners could potentially do more harm than good. 
This includes the risk of exploiting marginalized groups for our own career gain; Jessie states that: \textit{"I don't want to tokenize marginalized folks as a `selling point' for my research."} 
While significant funding can go towards quantitative research, if the marginalized communities are simply a vessel of studying the quantitative methods, the money most often could have been used more effectively by centering the marginalized group instead of the quantitative research. 

As discussed in Section \ref{sec:lessons-learned-from-pract}, researchers and their academic institutions are not solely focused on the objective of addressing the needs of marginalized communities. Researchers in departments that lack support and an existing structure for community-oriented work face particular challenges. One junior faculty member on our team mentions:
\begin{quote} 
\textit{
"It is difficult to balance your job requirements and dedicating the time to work with marginalized groups. The system is not set up to support you in this way, and if this is what you want to do, it requires a tremendous amount of effort to do so, especially if you are not in a community that is built to support this kind of collaboration."}
\end{quote}
Ultimately, researchers can successfully collaborate with practitioners as long as experienced guidance and support from superiors are given. Per Kristen, \textit{"the department or university or PI or whoever is applying high-level pressures about the outcome need to be on board with the needs of these kinds of projects."}

\subsection{Potential roles for researchers} 
How can researchers best support practitioners? We offer four suggestions: providing administrative tech support, acting as translators between different sectors, enhancing education through collaboration, and addressing the root causes of social issues.

\paragraph{Administrative tech support as volunteers}  
Quantitative researchers have an array of technical skills that can be a valuable contribution to the operations of community organizations. As Jessie pointed out, \textit{``Most people don’t need ML; they just need Excel.''} Further, Jessie stated that she considers how she can contribute, not in a role as researcher, but as \textit{"a member of society."} Similarly, others reflected that volunteering with organizations is one of the best options to make a concrete impact with our skill sets. In the future, we could take tech support roles as volunteers or collaborate on research with the practitioners. We highlight that these roles are non-exhaustive and are an option for people wishing to engage with and contribute to community advocacy groups. For larger, more involved projects under the umbrella of research, we see a role for focusing on technologies for usable administrative support, including digitization, databases, and tools for streamlining complex bureaucratic processes. We highlight that starting with engagement in administrative processes (e.g., helping create needed Excel documents) can be a first step in identifying further opportunities for useful technical interventions and high-impact research.

\paragraph{Translators} Researchers with experience working between the worlds of technology and community advocacy could actively seek opportunities as an intermediary between academia, the tech sector, the policy sphere, and non-profits. Rhetoric and jargon can often lead to misunderstandings when people with different expertise try to work together and learn from one another. Researchers can learn the language of practitioners and other domains when they talk and work with practitioners. The ability to communicate well through writing and reach a variety of audiences, especially those who are impacted by researchers' research, is incredibly valuable.

\paragraph{Education} The growth of AI, algorithms, and mechanism design in social sectors have sparked discussions about incorporating these applications into classroom teaching~\citep{lusk2022public}. Our proposed conversations with practitioners can contribute to more context-relevant teaching materials, such as case studies and course projects created through collaboration with practitioners.

Researchers can integrate practitioners' real-world challenges and context into course teaching in various forms. For instance, they could invite practitioners to give guest lectures as part of a course or collaboratively develop course projects or problem sets based on practitioners' contexts of applying AI in their work.\footnote{These teaching methods receive more attention in development economics and operations research, particularly in business schools. This supports the suggestion above that teaching methods modifications should suit each subject's specific context and teaching needs.} Academic researchers can conduct more comprehensive studies on practitioners' contexts to write case studies and incorporate them into their teaching. To prepare case studies, researchers must conduct in-depth interviews with practitioners, AI scholars, scientists, and engineers. These interviews can help researchers better understand the social contexts and cross-disciplinary perspectives to better frame technology problems of social applications that share similar features. 

Conversations with practitioners also help researchers enhance the short-term training programs and workshops for practitioners seeking to learn about AI, algorithms, and mechanism design technology. With a better understanding of the contexts different from where many theoretical frameworks and AI technology are developed, researchers can improve the curriculum and course design of training programs for practitioners.

\paragraph{Addressing the root causes} We recognize that some technology-focused projects aim, in the words of the current conference's mandate, to "help improve equity and access to opportunity for historically disadvantaged and under-served communities." \footnote{\url{https://www.eaamo.org/about}} Projects that focus on technical interventions to address structural inequities often have strong advocacy perspectives, from using data to push for policy changes to building tools serving the specific needs of under-served communities. Machine learning technologies can assist in collecting and organizing data on human rights abuses across the globe~\citep{huridocs2024HURIDOCSHumanRights}, including creating databases of gender-based violence from judicial documents or from news stories ~\citep{suresh2022IntersectionalFeministParticipatory}. Williams~\citep{dair2022DAIRFirstAnniversary} is creating a wage-theft calculator for Amazon delivery drivers to assist in advocating for changes to policies and regulations. To avoid data-extractivism or tokenization in the advancement of our research goals, our group must engage fully with the context and struggles of the groups with whom we work~\citep{lehuede2024DoubleHelixData}. Thus, we must engage with the root causes of complex social situations, often relating to issues of unbalanced power relations and systemic and historical discrimination.

\section{Concluding remarks}

We find that having a supportive environment to engage in interviews with practitioners has genuinely shifted our perspectives on doing research that engages with social issues. While we highlight many challenges and risks here, ultimately, each challenge and risk is counterbalanced by exciting opportunities and directions we had not considered before.

\begin{acks}
We are grateful to all the interviewees who shared their time with us and to our fellow group members in the Conversations with Practitioners working group at the Equity and Access in Algorithms, Mechanisms, and Optimization (EAAMO) initiative.

Mayra Russo and Kristen M. Scott are supported by the European Union's Horizon 2020 research and innovation program under Marie Sklodowska-Curie Actions (grant agreement number 860630) for the project "NoBIAS - Artificial Intelligence without Bias." This work reflects only the authors' views and the European Research Executive Agency (REA) is not responsible for any use that may be made of the information it contains.
This research received funding from the Flemish Government (AI Research Program).
Mackenzie Jorgensen is supported by the U.K. Research and Innovation under Grant EP/S023356/1 in the UKRI Centre for Doctoral Training in Safe and Trusted Artificial Intelligence.
Jessie Finocchiaro is supported by the National Science Foundation under Award No. 2202898.

\end{acks}

\bibliographystyle{ACM-Reference-Format}
\bibliography{references}

\appendix
\section{Autoethnographic Questions}
\label{sec:appendix}

\begin{enumerate}
\item What made you interested in joining the group?
\item Why did you become interested in interviews as a method?
\item What challenges did you face in organizing your interviews? What successes did you have, and what lessons did you learn?
\item How has this group impacted your thinking about working with practitioners and/or marginalized groups?
\item How has the group impacted your actual research?
\item What do you think about qualitative research? About the prospect of using it further in your own work?
\item What have you learned about the realities of working with marginalized groups in the current context in which you do research?
\item As a group, we have identified four specific 'lessons' that we have taken away from the experience
\begin{itemize}
\item Embrace the differing objectives of research and practice
\item Remember that quantification doesn't capture everything
\item Value the knowledge of experts
\item Avoid data-extractivism
\end{itemize}
Do these still resonate with you as lessons learned? Do others come to mind? What are your thoughts or experiences on implementing these lessons in your current or future work?
\end{enumerate}

\end{document}